\documentclass[preprint
  ,prd
  ,tightenlines
  ,superscriptaddress
]{revtex4-1}

\usepackage{graphicx} 
\usepackage{dcolumn}  
\usepackage{colordvi}
\usepackage{color}
\usepackage{pstricks}
\usepackage{epstopdf}
\usepackage{amssymb}
\usepackage{url}
\graphicspath{{ps}}
\usepackage{tabularx}
\usepackage{multirow}
\usepackage{units}
\usepackage{siunitx}
\usepackage{hyphenat}
\usepackage{subfigure}
\usepackage{float}
\usepackage[italic]{hepnames}
\usepackage{color, colortbl}

\renewcommand{\PBzero}{\ensuremath{\HepParticle{\PB}{}{}^0}\xspace}
\renewcommand{\APBzero}{\ensuremath{\HepParticle{\APB}{}{}^0}\xspace}

\renewcommand{\APDzero}{\ensuremath{\HepParticle{\APD}{}{}^0}\xspace}

\newcommand{\Bpipi}{\ensuremath{\APBzero \to \pi^0 \pi^0}\xspace}

\input{belle2sym}

\begin{document}

\def\belletwo {\it {Belle II}}

\vspace*{-3\baselineskip}
\resizebox{!}{3cm}{\includegraphics{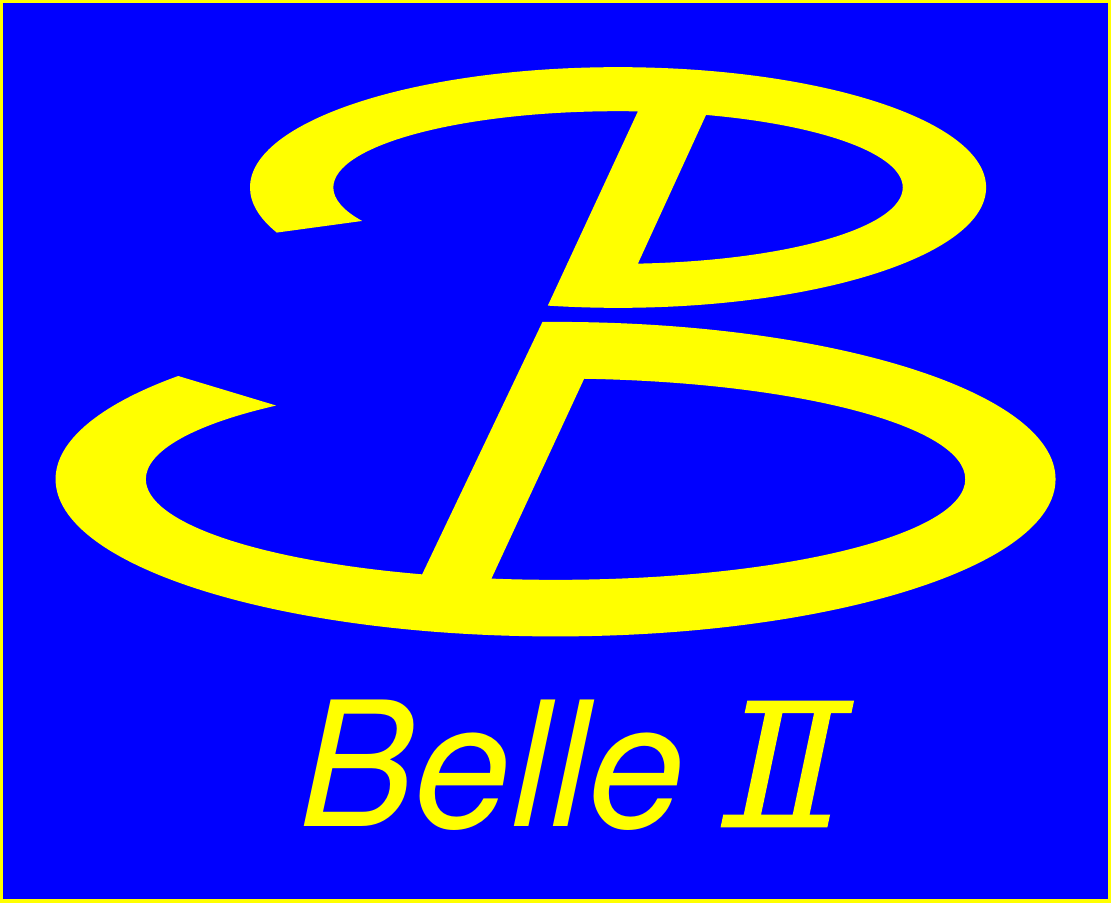}}

\vspace*{-5\baselineskip}
\begin{flushright}
BELLE2-CONF-PH-2021-010 \\
July 20, 2021

\end{flushright}

\vspace{1.5cm}
\title {Measurement of the branching fraction for $B^{0} \to \pi^{0} \pi^{0}$ decays reconstructed in 2019–2020 Belle II data}

\newcommand{\instCPPM}{Aix Marseille Universit\'{e}, CNRS/IN2P3, CPPM, 13288 Marseille, France}
\newcommand{\instBeihang}{Beihang University, Beijing 100191, China}
\newcommand{\instBNL}{Brookhaven National Laboratory, Upton, New York 11973, U.S.A.}
\newcommand{\instBINP}{Budker Institute of Nuclear Physics SB RAS, Novosibirsk 630090, Russian Federation}
\newcommand{\instCMU}{Carnegie Mellon University, Pittsburgh, Pennsylvania 15213, U.S.A.}
\newcommand{\instCinvestavIPN}{Centro de Investigacion y de Estudios Avanzados del Instituto Politecnico Nacional, Mexico City 07360, Mexico}
\newcommand{\instPrague}{Faculty of Mathematics and Physics, Charles University, 121 16 Prague, Czech Republic}
\newcommand{\instChiangMai}{Chiang Mai University, Chiang Mai 50202, Thailand}
\newcommand{\instChiba}{Chiba University, Chiba 263-8522, Japan}
\newcommand{\instChonnam}{Chonnam National University, Gwangju 61186, South Korea}
\newcommand{\instConacyt}{Consejo Nacional de Ciencia y Tecnolog\'{\i}a, Mexico City 03940, Mexico}
\newcommand{\instDESY}{Deutsches Elektronen--Synchrotron, 22607 Hamburg, Germany}
\newcommand{\instDuke}{Duke University, Durham, North Carolina 27708, U.S.A.}
\newcommand{\instITAR}{Institute of Theoretical and Applied Research (ITAR), Duy Tan University, Hanoi 100000, Vietnam}
\newcommand{\instRomaENEA}{ENEA Casaccia, I-00123 Roma, Italy}
\newcommand{\instEri}{Earthquake Research Institute, University of Tokyo, Tokyo 113-0032, Japan}
\newcommand{\instJuelich}{Forschungszentrum J\"{u}lich, 52425 J\"{u}lich, Germany}
\newcommand{\instFuJen}{Department of Physics, Fu Jen Catholic University, Taipei 24205, Taiwan}
\newcommand{\instFudan}{Key Laboratory of Nuclear Physics and Ion-beam Application (MOE) and Institute of Modern Physics, Fudan University, Shanghai 200443, China}
\newcommand{\instGoettingen}{II. Physikalisches Institut, Georg-August-Universit\"{a}t G\"{o}ttingen, 37073 G\"{o}ttingen, Germany}
\newcommand{\instGifu}{Gifu University, Gifu 501-1193, Japan}
\newcommand{\instSOKENDAI}{The Graduate University for Advanced Studies (SOKENDAI), Hayama 240-0193, Japan}
\newcommand{\instGyeongsang}{Gyeongsang National University, Jinju 52828, South Korea}
\newcommand{\instHanyang}{Department of Physics and Institute of Natural Sciences, Hanyang University, Seoul 04763, South Korea}
\newcommand{\instKEK}{High Energy Accelerator Research Organization (KEK), Tsukuba 305-0801, Japan}
\newcommand{\instJPARC}{J-PARC Branch, KEK Theory Center, High Energy Accelerator Research Organization (KEK), Tsukuba 305-0801, Japan}
\newcommand{\instHSE}{National Research University Higher School of Economics, Moscow 101000, Russian Federation}
\newcommand{\instIISER}{Indian Institute of Science Education and Research Mohali, SAS Nagar, 140306, India}
\newcommand{\instIITBhubaneswar}{Indian Institute of Technology Bhubaneswar, Satya Nagar 751007, India}
\newcommand{\instIITGuwahati}{Indian Institute of Technology Guwahati, Assam 781039, India}
\newcommand{\instIITHyderabad}{Indian Institute of Technology Hyderabad, Telangana 502285, India}
\newcommand{\instIITMadras}{Indian Institute of Technology Madras, Chennai 600036, India}
\newcommand{\instIndiana}{Indiana University, Bloomington, Indiana 47408, U.S.A.}
\newcommand{\instIHEPRussia}{Institute for High Energy Physics, Protvino 142281, Russian Federation}
\newcommand{\instHEPHYVienna}{Institute of High Energy Physics, Vienna 1050, Austria}
\newcommand{\instHiroshima}{Hiroshima University, Higashi-Hiroshima, Hiroshima 739-8530, Japan}
\newcommand{\instIHEPChina}{Institute of High Energy Physics, Chinese Academy of Sciences, Beijing 100049, China}
\newcommand{\instIPP}{Institute of Particle Physics (Canada), Victoria, British Columbia V8W 2Y2, Canada}
\newcommand{\instIOP}{Institute of Physics, Vietnam Academy of Science and Technology (VAST), Hanoi, Vietnam}
\newcommand{\instIFIC}{Instituto de Fisica Corpuscular, Paterna 46980, Spain}
\newcommand{\instFrascati}{INFN Laboratori Nazionali di Frascati, I-00044 Frascati, Italy}
\newcommand{\instNapoliINFN}{INFN Sezione di Napoli, I-80126 Napoli, Italy}
\newcommand{\instPadovaINFN}{INFN Sezione di Padova, I-35131 Padova, Italy}
\newcommand{\instPerugiaINFN}{INFN Sezione di Perugia, I-06123 Perugia, Italy}
\newcommand{\instPisaINFN}{INFN Sezione di Pisa, I-56127 Pisa, Italy}
\newcommand{\instRomaINFN}{INFN Sezione di Roma, I-00185 Roma, Italy}
\newcommand{\instRomaTreINFN}{INFN Sezione di Roma Tre, I-00146 Roma, Italy}
\newcommand{\instTorinoINFN}{INFN Sezione di Torino, I-10125 Torino, Italy}
\newcommand{\instTriesteINFN}{INFN Sezione di Trieste, I-34127 Trieste, Italy}
\newcommand{\instJAEA}{Advanced Science Research Center, Japan Atomic Energy Agency, Naka 319-1195, Japan}
\newcommand{\instMainz}{Johannes Gutenberg-Universit\"{a}t Mainz, Institut f\"{u}r Kernphysik, D-55099 Mainz, Germany}
\newcommand{\instGiessen}{Justus-Liebig-Universit\"{a}t Gie\ss{}en, 35392 Gie\ss{}en, Germany}
\newcommand{\instKarlsruhe}{Institut f\"{u}r Experimentelle Teilchenphysik, Karlsruher Institut f\"{u}r Technologie, 76131 Karlsruhe, Germany}
\newcommand{\instISU}{Iowa State University, Ames, Iowa 50011, U.S.A.}
\newcommand{\instKitasato}{Kitasato University, Sagamihara 252-0373, Japan}
\newcommand{\instKISTI}{Korea Institute of Science and Technology Information, Daejeon 34141, South Korea}
\newcommand{\instKoreaUnivKU}{Korea University, Seoul 02841, South Korea}
\newcommand{\instKSU}{Kyoto Sangyo University, Kyoto 603-8555, Japan}
\newcommand{\instKyungpook}{Kyungpook National University, Daegu 41566, South Korea}
\newcommand{\instLPI}{P.N. Lebedev Physical Institute of the Russian Academy of Sciences, Moscow 119991, Russian Federation}
\newcommand{\instLNNU}{Liaoning Normal University, Dalian 116029, China}
\newcommand{\instLMU}{Ludwig Maximilians University, 80539 Munich, Germany}
\newcommand{\instLuther}{Luther College, Decorah, Iowa 52101, U.S.A.}
\newcommand{\instMNITJaipur}{Malaviya National Institute of Technology Jaipur, Jaipur 302017, India}
\newcommand{\instMPP}{Max-Planck-Institut f\"{u}r Physik, 80805 M\"{u}nchen, Germany}
\newcommand{\instMPGHLL}{Semiconductor Laboratory of the Max Planck Society, 81739 M\"{u}nchen, Germany}
\newcommand{\instMcGill}{McGill University, Montr\'{e}al, Qu\'{e}bec, H3A 2T8, Canada}
\newcommand{\instMEPhI}{Moscow Physical Engineering Institute, Moscow 115409, Russian Federation}
\newcommand{\instNagoya}{Graduate School of Science, Nagoya University, Nagoya 464-8602, Japan}
\newcommand{\instNagoyaIAR}{Institute for Advanced Research, Nagoya University, Nagoya 464-8602, Japan}
\newcommand{\instNagoyaKMI}{Kobayashi-Maskawa Institute, Nagoya University, Nagoya 464-8602, Japan}
\newcommand{\instNaraWu}{Nara Women's University, Nara 630-8506, Japan}
\newcommand{\instNTUTaiwan}{Department of Physics, National Taiwan University, Taipei 10617, Taiwan}
\newcommand{\instNUUTaiwan}{National United University, Miao Li 36003, Taiwan}
\newcommand{\instKrakow}{H. Niewodniczanski Institute of Nuclear Physics, Krakow 31-342, Poland}
\newcommand{\instNiigata}{Niigata University, Niigata 950-2181, Japan}
\newcommand{\instNSU}{Novosibirsk State University, Novosibirsk 630090, Russian Federation}
\newcommand{\instOkinawa}{Okinawa Institute of Science and Technology, Okinawa 904-0495, Japan}
\newcommand{\instOsakaCity}{Osaka City University, Osaka 558-8585, Japan}
\newcommand{\instRCNP}{Research Center for Nuclear Physics, Osaka University, Osaka 567-0047, Japan}
\newcommand{\instPNNL}{Pacific Northwest National Laboratory, Richland, Washington 99352, U.S.A.}
\newcommand{\instPanjab}{Panjab University, Chandigarh 160014, India}
\newcommand{\instPanjabPAU}{Punjab Agricultural University, Ludhiana 141004, India}
\newcommand{\instRIKENMSL}{Meson Science Laboratory, Cluster for Pioneering Research, RIKEN, Saitama 351-0198, Japan}
\newcommand{\instSeoul}{Seoul National University, Seoul 08826, South Korea}
\newcommand{\instSPU}{Showa Pharmaceutical University, Tokyo 194-8543, Japan}
\newcommand{\instSoochow}{Soochow University, Suzhou 215006, China}
\newcommand{\instSoongsil}{Soongsil University, Seoul 06978, South Korea}
\newcommand{\instLjubljanaJSI}{J. Stefan Institute, 1000 Ljubljana, Slovenia}
\newcommand{\instKyiv}{Taras Shevchenko National Univ. of Kiev, Kiev, Ukraine}
\newcommand{\instTata}{Tata Institute of Fundamental Research, Mumbai 400005, India}
\newcommand{\instTUM}{Department of Physics, Technische Universit\"{a}t M\"{u}nchen, 85748 Garching, Germany}
\newcommand{\instTelAviv}{Tel Aviv University, School of Physics and Astronomy, Tel Aviv, 69978, Israel}
\newcommand{\instToho}{Toho University, Funabashi 274-8510, Japan}
\newcommand{\instTohoku}{Department of Physics, Tohoku University, Sendai 980-8578, Japan}
\newcommand{\instTitech}{Tokyo Institute of Technology, Tokyo 152-8550, Japan}
\newcommand{\instTokyoMetropolitan}{Tokyo Metropolitan University, Tokyo 192-0397, Japan}
\newcommand{\instUAS}{Universidad Autonoma de Sinaloa, Sinaloa 80000, Mexico}
\newcommand{\instNapoliUNIV}{Dipartimento di Scienze Fisiche, Universit\`{a} di Napoli Federico II, I-80126 Napoli, Italy}
\newcommand{\instPadovaUNIV}{Dipartimento di Fisica e Astronomia, Universit\`{a} di Padova, I-35131 Padova, Italy}
\newcommand{\instPerugiaUNIV}{Dipartimento di Fisica, Universit\`{a} di Perugia, I-06123 Perugia, Italy}
\newcommand{\instPisaUNIV}{Dipartimento di Fisica, Universit\`{a} di Pisa, I-56127 Pisa, Italy}
\newcommand{\instRomaTreUNIV}{Dipartimento di Matematica e Fisica, Universit\`{a} di Roma Tre, I-00146 Roma, Italy}
\newcommand{\instTorinoUNIV}{Dipartimento di Fisica, Universit\`{a} di Torino, I-10125 Torino, Italy}
\newcommand{\instTriesteUNIV}{Dipartimento di Fisica, Universit\`{a} di Trieste, I-34127 Trieste, Italy}
\newcommand{\instMontreal}{Universit\'{e} de Montr\'{e}al, Physique des Particules, Montr\'{e}al, Qu\'{e}bec, H3C 3J7, Canada}
\newcommand{\instIJCLab}{Universit\'{e} Paris-Saclay, CNRS/IN2P3, IJCLab, 91405 Orsay, France}
\newcommand{\instIPHC}{Universit\'{e} de Strasbourg, CNRS, IPHC, UMR 7178, 67037 Strasbourg, France}
\newcommand{\instAdelaide}{Department of Physics, University of Adelaide, Adelaide, South Australia 5005, Australia}
\newcommand{\instBonn}{University of Bonn, 53115 Bonn, Germany}
\newcommand{\instUBC}{University of British Columbia, Vancouver, British Columbia, V6T 1Z1, Canada}
\newcommand{\instCincinnati}{University of Cincinnati, Cincinnati, Ohio 45221, U.S.A.}
\newcommand{\instFlorida}{University of Florida, Gainesville, Florida 32611, U.S.A.}
\newcommand{\instHawaii}{University of Hawaii, Honolulu, Hawaii 96822, U.S.A.}
\newcommand{\instHeidelberg}{University of Heidelberg, 68131 Mannheim, Germany}
\newcommand{\instLjubljanaUniLJ}{Faculty of Mathematics and Physics, University of Ljubljana, 1000 Ljubljana, Slovenia}
\newcommand{\instLouisville}{University of Louisville, Louisville, Kentucky 40292, U.S.A.}
\newcommand{\instMalaya}{National Centre for Particle Physics, University Malaya, 50603 Kuala Lumpur, Malaysia}
\newcommand{\instLjubljanaUM}{Faculty of Chemistry and Chemical Engineering, University of Maribor, 2000 Maribor, Slovenia}
\newcommand{\instMelbourne}{School of Physics, University of Melbourne, Victoria 3010, Australia}
\newcommand{\instMississippi}{University of Mississippi, University, Mississippi 38677, U.S.A.}
\newcommand{\instUOM}{University of Miyazaki, Miyazaki 889-2192, Japan}
\newcommand{\instPittsburgh}{University of Pittsburgh, Pittsburgh, Pennsylvania 15260, U.S.A.}
\newcommand{\instUSTC}{University of Science and Technology of China, Hefei 230026, China}
\newcommand{\instSAlabama}{University of South Alabama, Mobile, Alabama 36688, U.S.A.}
\newcommand{\instSCarolina}{University of South Carolina, Columbia, South Carolina 29208, U.S.A.}
\newcommand{\instSydney}{School of Physics, University of Sydney, New South Wales 2006, Australia}
\newcommand{\instUTokyo}{Department of Physics, University of Tokyo, Tokyo 113-0033, Japan}
\newcommand{\instIPMU}{Kavli Institute for the Physics and Mathematics of the Universe (WPI), University of Tokyo, Kashiwa 277-8583, Japan}
\newcommand{\instVictoria}{University of Victoria, Victoria, British Columbia, V8W 3P6, Canada}
\newcommand{\instVPI}{Virginia Polytechnic Institute and State University, Blacksburg, Virginia 24061, U.S.A.}
\newcommand{\instWayneState}{Wayne State University, Detroit, Michigan 48202, U.S.A.}
\newcommand{\instYamagata}{Yamagata University, Yamagata 990-8560, Japan}
\newcommand{\instYerevan}{Alikhanyan National Science Laboratory, Yerevan 0036, Armenia}
\newcommand{\instYonsei}{Yonsei University, Seoul 03722, South Korea}
\newcommand{\instZZU}{Zhengzhou University, Zhengzhou 450001, China}
\affiliation{\instCPPM}
\affiliation{\instBeihang}
\affiliation{\instBNL}
\affiliation{\instBINP}
\affiliation{\instCMU}
\affiliation{\instCinvestavIPN}
\affiliation{\instPrague}
\affiliation{\instChiangMai}
\affiliation{\instChiba}
\affiliation{\instChonnam}
\affiliation{\instConacyt}
\affiliation{\instDESY}
\affiliation{\instDuke}
\affiliation{\instITAR}
\affiliation{\instRomaENEA}
\affiliation{\instEri}
\affiliation{\instJuelich}
\affiliation{\instFuJen}
\affiliation{\instFudan}
\affiliation{\instGoettingen}
\affiliation{\instGifu}
\affiliation{\instSOKENDAI}
\affiliation{\instGyeongsang}
\affiliation{\instHanyang}
\affiliation{\instKEK}
\affiliation{\instJPARC}
\affiliation{\instHSE}
\affiliation{\instIISER}
\affiliation{\instIITBhubaneswar}
\affiliation{\instIITGuwahati}
\affiliation{\instIITHyderabad}
\affiliation{\instIITMadras}
\affiliation{\instIndiana}
\affiliation{\instIHEPRussia}
\affiliation{\instHEPHYVienna}
\affiliation{\instHiroshima}
\affiliation{\instIHEPChina}
\affiliation{\instIPP}
\affiliation{\instIOP}
\affiliation{\instIFIC}
\affiliation{\instFrascati}
\affiliation{\instNapoliINFN}
\affiliation{\instPadovaINFN}
\affiliation{\instPerugiaINFN}
\affiliation{\instPisaINFN}
\affiliation{\instRomaINFN}
\affiliation{\instRomaTreINFN}
\affiliation{\instTorinoINFN}
\affiliation{\instTriesteINFN}
\affiliation{\instJAEA}
\affiliation{\instMainz}
\affiliation{\instGiessen}
\affiliation{\instKarlsruhe}
\affiliation{\instISU}
\affiliation{\instKitasato}
\affiliation{\instKISTI}
\affiliation{\instKoreaUnivKU}
\affiliation{\instKSU}
\affiliation{\instKyungpook}
\affiliation{\instLPI}
\affiliation{\instLNNU}
\affiliation{\instLMU}
\affiliation{\instLuther}
\affiliation{\instMNITJaipur}
\affiliation{\instMPP}
\affiliation{\instMPGHLL}
\affiliation{\instMcGill}
\affiliation{\instMEPhI}
\affiliation{\instNagoya}
\affiliation{\instNagoyaIAR}
\affiliation{\instNagoyaKMI}
\affiliation{\instNaraWu}
\affiliation{\instNTUTaiwan}
\affiliation{\instNUUTaiwan}
\affiliation{\instKrakow}
\affiliation{\instNiigata}
\affiliation{\instNSU}
\affiliation{\instOkinawa}
\affiliation{\instOsakaCity}
\affiliation{\instRCNP}
\affiliation{\instPNNL}
\affiliation{\instPanjab}
\affiliation{\instPanjabPAU}
\affiliation{\instRIKENMSL}
\affiliation{\instSeoul}
\affiliation{\instSPU}
\affiliation{\instSoochow}
\affiliation{\instSoongsil}
\affiliation{\instLjubljanaJSI}
\affiliation{\instKyiv}
\affiliation{\instTata}
\affiliation{\instTUM}
\affiliation{\instTelAviv}
\affiliation{\instToho}
\affiliation{\instTohoku}
\affiliation{\instTitech}
\affiliation{\instTokyoMetropolitan}
\affiliation{\instUAS}
\affiliation{\instNapoliUNIV}
\affiliation{\instPadovaUNIV}
\affiliation{\instPerugiaUNIV}
\affiliation{\instPisaUNIV}
\affiliation{\instRomaTreUNIV}
\affiliation{\instTorinoUNIV}
\affiliation{\instTriesteUNIV}
\affiliation{\instMontreal}
\affiliation{\instIJCLab}
\affiliation{\instIPHC}
\affiliation{\instAdelaide}
\affiliation{\instBonn}
\affiliation{\instUBC}
\affiliation{\instCincinnati}
\affiliation{\instFlorida}
\affiliation{\instHawaii}
\affiliation{\instHeidelberg}
\affiliation{\instLjubljanaUniLJ}
\affiliation{\instLouisville}
\affiliation{\instMalaya}
\affiliation{\instLjubljanaUM}
\affiliation{\instMelbourne}
\affiliation{\instMississippi}
\affiliation{\instUOM}
\affiliation{\instPittsburgh}
\affiliation{\instUSTC}
\affiliation{\instSAlabama}
\affiliation{\instSCarolina}
\affiliation{\instSydney}
\affiliation{\instUTokyo}
\affiliation{\instIPMU}
\affiliation{\instVictoria}
\affiliation{\instVPI}
\affiliation{\instWayneState}
\affiliation{\instYamagata}
\affiliation{\instYerevan}
\affiliation{\instYonsei}
\affiliation{\instZZU}
  \author{F.~Abudin{\'e}n}\affiliation{\instTriesteINFN} 
  \author{I.~Adachi}\affiliation{\instKEK}\affiliation{\instSOKENDAI} 
  \author{R.~Adak}\affiliation{\instFudan} 
  \author{K.~Adamczyk}\affiliation{\instKrakow} 
  \author{P.~Ahlburg}\affiliation{\instBonn} 
  \author{J.~K.~Ahn}\affiliation{\instKoreaUnivKU} 
  \author{H.~Aihara}\affiliation{\instUTokyo} 
  \author{N.~Akopov}\affiliation{\instYerevan} 
  \author{A.~Aloisio}\affiliation{\instNapoliUNIV}\affiliation{\instNapoliINFN} 
  \author{F.~Ameli}\affiliation{\instRomaINFN} 
% ...

\begin{abstract}
We report the first reconstruction of the $B^{0} \to \pi^{0} \pi^{0}$ decay mode at Belle II using samples of 2019 and 2020 data that correspond to 62.8 \invfb of integrated luminosity. We find $14.0^{+6.8}_{-5.6}$ signal decays, corresponding to a significance of 3.4 standard deviations and determine a branching ratio of $\mathcal{B}(\Bpipi) = [0.98^{+0.48}_{-0.39} \pm 0.27] \times 10^{-6}$. The results agree with previous determinations and contribute important information to an early assessment of detector performance and Belle II's potential for future determinations of $\alpha/\phi_2$ using $\PB \to \pi \pi$ modes.

\end{abstract}


\maketitle

{\renewcommand{\thefootnote}{\fnsymbol{footnote}}}
\setcounter{footnote}{0}


\newpage

\section{Introduction}

\begin{figure}[!h]
    \centering
    \includegraphics[width=0.5\textwidth]{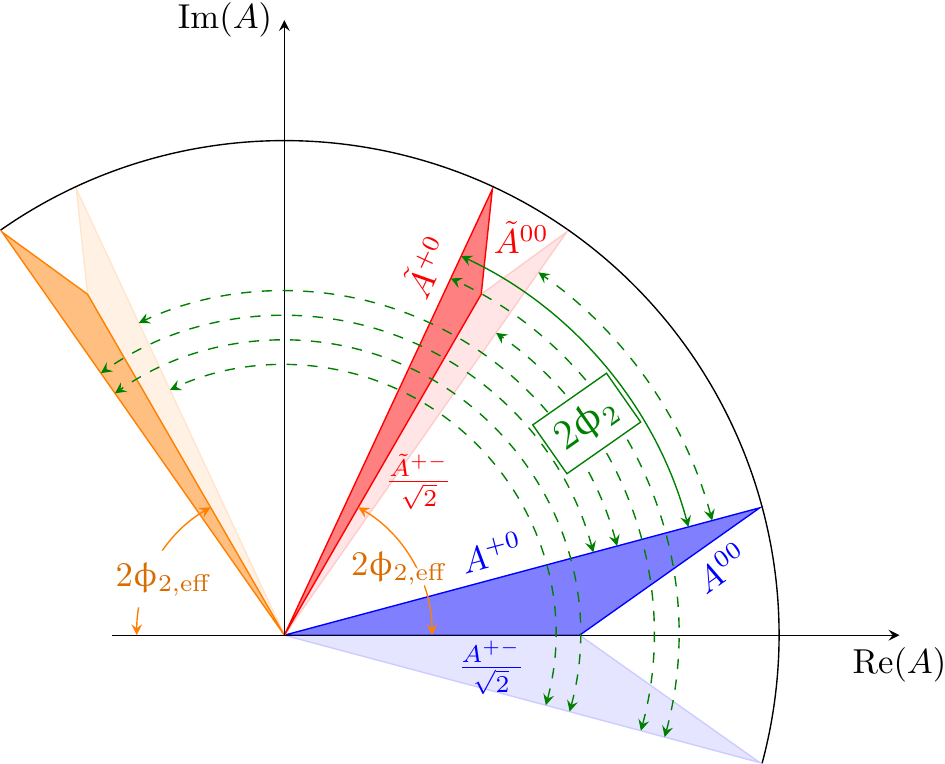}
    \caption{Geometrical representation of the isospin triangular relations in the complex plane of $\PB^{i+j}\to \Ph^i\Ph^j$ amplitudes. The blue and the red shaded areas correspond to the isospin triangles. The angle between the ${\it CP}$ conjugate charged amplitudes   $A^{+-}$ and $\bar{A}^{+-}$ corresponds to twice the weak phase $\alpha_{\rm eff}/\phi_{2,{\rm eff}}$ (orange solid arcs). The angle between the ${\it CP}$ conjugate charged amplitudes $A^{+0}$ and $\bar{A}^{+0}$ corresponds to twice the CKM angle $\alpha/\phi_{2}$ (green solid arc). The other triangles with lighter shading represent the mirror solutions allowed by the discrete ambiguities in the isospin relationships, with the corresponding values for $\alpha/\phi_{2}$  represented by the green dashed curves.}
    \label{fig:isospin}
\end{figure}

Effective constraints on physics beyond the Standard Model are provided by high-precision measurements. The study of charmless decays at Belle II can provide improved measurements of the Cabibbo–Kobayashi–Maskawa (CKM) angle $\alpha/\phi_2 \equiv \textrm{arg}( -\frac{V_{td} V^*_{cb}}{V_{ud} V^*_{ub}} )$, where $V_{ij}$ are elements of the quark-mixing matrix. If $\PBzero \rightarrow \pi^+ \pi^-$ decays proceeded through only the tree level ($b \rightarrow u$) process, the mixing-induced ${\it CP}$ violation parameter, $\mathcal{S}_{CP}$, would be proportional to sin$(2\phi_2)$. However the value of $\phi_2$ is shifted by an amount $\Delta \phi_2$ due to the presence of penguin contributions ($b \rightarrow d$). The tree and penguin contributions can be disentangled using $B \rightarrow \pi \pi$ isospin relations~\cite{Gronau1990}.
\begin{equation}
     A^{+0} = \frac{1}{\sqrt{2}}A^{+-} + A^{00}, \quad \bar{A}^{-0} = \frac{1}{\sqrt{2}}\bar{A}^{+-} + \bar{A}^{00}
\end{equation}
where $A^{ij}$ is the amplitude of the decay $\bar{B} \to \pi^i \pi^j$ and is represented geometrically in the complex plane in Figure \ref{fig:isospin}. 

Taking advantage of these relations requires precise measurements of the branching fraction $\mathcal{B}$ and ${\it CP}$ violation parameters of each $B \rightarrow \pi \pi$ decay. The relatively large uncertainties on the current value of $\mathcal{B}(\PBzero \rightarrow \pi^0 \pi^0)$ and $\mathcal{A}_{CP}(\PBzero \rightarrow \pi^0 \pi^0)$, measured by BaBar~\cite{BaBar:2007uoe} and Belle~\cite{Belle:2017lyb}, poses the greatest limitation to fully exploiting the isospin relation. The $\PBzero \rightarrow \pi^0 \pi^0$ mode has a low branching ratio, ($1.59 \pm 0.26) \times 10^{-6}$~\cite{Zyla:2020zbs}, since it is both color-suppressed and, at tree level, is proportional to the CKM matrix element $V_{ub}$, whose magnitude is small. In addition, the $\pi^0$ decays via $\pi^0 \to \gamma \gamma$ with a branching ratio of $\approx$ 99\% and hence the final state particles consist entirely of photons. Belle II with its clean $e^+e^-$ environment and large acceptance for photons ranging from 20 MeV to 4 GeV is the only running experiment that can competitively study this decay mode.  

The Belle II experiment features significantly upgraded detectors and new analysis software providing better particle identification, background suppression and $\it B$-meson flavour determination compared to its predecessor Belle. The Belle II experiment, complete with its silicon vertex detector, commenced taking data in March 2019. The sample used in this analysis corresponds to an integrated luminosity of 62.8 \invfb at the $\Upsilon(4S)$ resonance. We report the first Belle II measurement of the branching fraction of the $B^{0} \to \pi^{0} \pi^{0}$ decay. Charge-conjugate decays are implied in what follows.

\section{Belle II detector}

A full description of the Belle II detector is given in Ref.~\cite{Belle-II:2010dht}. The detector consists of several subdetectors arranged in a cylindrical structure around the beam pipe. Compared to its predecessor Belle~\cite{Belle:2000cnh}, a pixel detector (PXD) has been added at a minimum radius of 1.4 cm. This improves the resolution of the impact parameter to about 12 $\mu$m in the transverse direction for high momentum tracks~\cite{Belle-II:2018jsg}, which helps to reject background events for this analysis. The PXD is surrounded by a four-layer double-sided silicon strip detector, referred to as the silicon vertex detector, and a central drift chamber (CDC). A time-of-propagation counter and an aerogel ring-imaging Cherenkov counter cover the barrel and forward endcap regions of the detector, respectively, and are essential for charged-particle identification (PID). The electromagnetic calorimeter (ECL) makes up the remaining volume inside a superconducting solenoid, which operates at 1.5 T. A dedicated detector to identify $K^0_L$ mesons and muons is installed in the outermost part of the detector. The z-axis of the laboratory frame is defined as the symmetry axis of the solenoid, and the positive direction is approximately given by the incoming electron beam. The polar angle $\theta$, as well as the longitudinal and the transverse direction are defined with respect to the z-axis. The ECL is most relevant for this work as it is the only subdetector that can detect photons.

\subsection{Data and simulation}

We use all 2019-2020 data corresponding to an integrated luminosity of 62.8 \invfb collected with the Belle II detector at the asymmetric-energy $\epem$ collider SuperKEKB~\cite{Akai:2018mbz}, which is located at the KEK laboratory in Tsukuba, Japan. Data were collected at the center-of-mass (c.m.) energy of the \Y4S resonance ($\sqrt{s}=  10.58 \gev$). The energies of the electron and positron beams are 7 GeV and 4 GeV, respectively, resulting in a boost of $\beta\gamma = 0.28$ of the c.m. frame relative to the laboratory frame. We also use all off-resonance data collected at an energy about 60 MeV lower and corresponds to an integrated luminosity of 9.2 \invfb. All events are required to satisfy loose hadronic event selection criteria, based on total energy and neutral-particle multiplicity in the events, targeted at reducing sample sizes to a manageable level with minimal impact on signal efficiency. All data are processed using the Belle II analysis software~\cite{Kuhr:2018lps}. 

We use GEANT4~\cite{GEANT4:2002zbu} based Monte Carlo~(MC) simulation data to optimize the event selection, compare the distribution observed in experimental data with expectations, and model the distribution in fits. We use $2\times 10^{6}$ signal simulated data. Generic background MC samples consists of charged and neutral B meson pairs ($\PBzero \APBzero$ and $B^{+} B^{-}$), and continuum processes ($e^+e^- \to$ \qqbar with $q = u, d ,s ,c$ quarks) in realistic proportions, which correspond to a 4 \invab sample. To validate our experimental procedure, the $B^{0} \rightarrow \APDzero(\rightarrow K^- \pi^+ \pi^0) \pi^0$ decay is used as a control mode as it contains two $\pi^0$ particles in the final state. The total yield is expected to be 10 -- 20 times larger than the expected signal yield from the $B^{0} \to \pi^{0} \pi^{0}$ decay. We use $1 \times 10^{6}$ simulated control-mode events.

\section{Event Selection and Candidate Reconstruction}
\label{sec:selection}

We form photon candidates by requiring the energy in the ECL barrel and endcaps to be greater than 20 MeV and 22.5 MeV respectively. Further photon selections using a binary boosted decision-tree classifier, described in subsection 3.1, is applied to suppress hadronic interactions and photons from non-signal sources. The photon candidates are paired to form $\pi^0$ candidates and we require the invariant mass and helicitiy angle to be $0.105 < M_{\gamma \gamma} < 0.150 \gevcc$ and $|\text{cos} \theta_{helicity}| < 0.98$, respectively, to suppress combinatorial background from collinear soft photons. The mass of the \piz candidates is constrained to its known value in a kinematic fit to improve the momentum resolution. The $B$ meson candidates are reconstructed by pairing the $\pi^0$ candidates. To select signal $B$, two kinematic variables are defined, 

\begin{equation}
    \mbc = \sqrt{E_{\rm beam}^2 - |\vec{p}_{B}|^2}, \quad \dele = E_{B} - \sqrt{s}/2 
\end{equation}

where \mbc, is the mass-energy relation where the energy of the $B$ meson has been replaced by half of the c.m. energy, which is extremely well-defined by the accelerator, and $\vec{p_{B}}$ is the $B$ meson momentum, both measured in the $\Upsilon(4S)$ frame. $\dele = E_{B} - \sqrt{s}/2$ is the difference between the total energy of the $B$ candidate and half the collision energy, both measured in the $\Upsilon(4S)$ frame. $B$ meson candidates are required to have $5.26 < \mbc < 5.29~\gevcc$ and $-0.3 < \dele < 0.2~\gev$. For correctly reconstructed $B$ meson candidates, $\dele$ should peak at zero except for resolution. However, observed $\dele$ distributions peak below zero since energy is lost via either electromagnetic interactions in the material before the calorimeter or via energy leakage from the ECL cluster. In addition, the photon energy and momentum can only be determined from the ECL cluster energy and hence there is a small correlation between $\dele$ and $\mbc$ in our reconstructed $\PBzero \rightarrow \pi^0 \pi^0$ events.

\subsection{Optimized photon selection} 
\label{photonmva}

Due to the long decay time of signal in CsI(Tl), Bhabha events ($e^+e^- \to e^+e^-$) can deposit large amounts of energy in the CsI(Tl) crystals of the ECL that are still present when another hadronic event occurs. A random photon from the hadronic event can be combined with the residual energy (misreconstructed photon) in the CsI(Tl) crystals to form a $\pi^0$. This misreconstructed $\pi^0$ and a genuine $\pi^0$ can then be misreconstructed into a $\PBzero$ candidate. In the Belle analysis of $\PBzero \rightarrow \pi^0 \pi^0$~\cite{Julius:2017jso}, these were suppressed by requiring ECL signals to be in time with the rest of the $e^{+}e^{-}$ event. Rather than using one-dimensional requirements on the photon's time of interaction, we train a fast boosted decision-tree (FBDT)~\cite{Keck:2017gsv} to distinguish between genuine and misreconstructed photons using ECL variables that have high discriminating power. To create samples of genuine and misreconstructed photons, we reconstruct $\Bpipi$ candidates in simulated signal-only data with no requirements on the photons. Photons that do not originate from the signal are regarded as misreconstructed. We train and validate data using two independent data sets consisting of 50~000 genuine and 50~000 misreconstructed photons. We find that this classifier, which we call the \texttt{photonMVA}, can efficiently suppress most misreconstructed photons. Based on MC studies requiring the \texttt{photonMVA} output to exceed 0.2 retains 97.05\% of genuine photons while rejecting 73.3\% of misreconstructed photons. 

To validate the \texttt{photonMVA} we use the $D^{*+} \rightarrow \APDzero(\rightarrow K^0_S (\rightarrow \pi^+ \pi^-) \pi^0) \pi^+$ decay mode. The reconstruction uses similar final state selections as the signal decay. We first reconstruct $\pi^{\pm}$ candidates from charged-particle candidates reconstructed in the full polar-angle acceptance ($17\degrees < \theta < 150\degrees$), and originating close to the interaction point in the longitudinal ($|dz|<3.0$ cm) and radial ($|dr|<0.5$ cm) directions to reduce beam-background-induced tracks. For ${K^0_s}$ reconstruction, we pair oppositely charged $\pi^{\pm}$ candidates and require that they originate from a common space-point and have dipion mass in the range 0.47 -- 0.53~\gevcc. For $D^{0}$ reconstruction, we combine the ${K^0_s}$ with a $\pi^0$ and require that candidates have masses in the range 1.80 -- 2.50~\gevcc with momenta in the c.m. frame greater than 2.5~\gevc. Finally, we reconstruct $D^{*+}$ candidates by combining the $D^{0}$ and $\pi^+$ in a kinematic vertex fit. We choose one candidate per event by selecting the $\pi^0$ candidate with the lowest $\chi^2$ value of the mass-constraint diphoton fit. The difference between the mass of the $D^{*+}$ and $\APDzero$, $\Delta M = M_{D^{*+}} - M_{\APDzero}$, is a powerful discriminator between signal and background. We require $0.144 < \Delta M < 0.147 \gevcc$ to ensure the $D^{*+}$ is reconstructed with high purity. 

\begin{figure}
    \centering
    \includegraphics[width=0.7\textwidth]{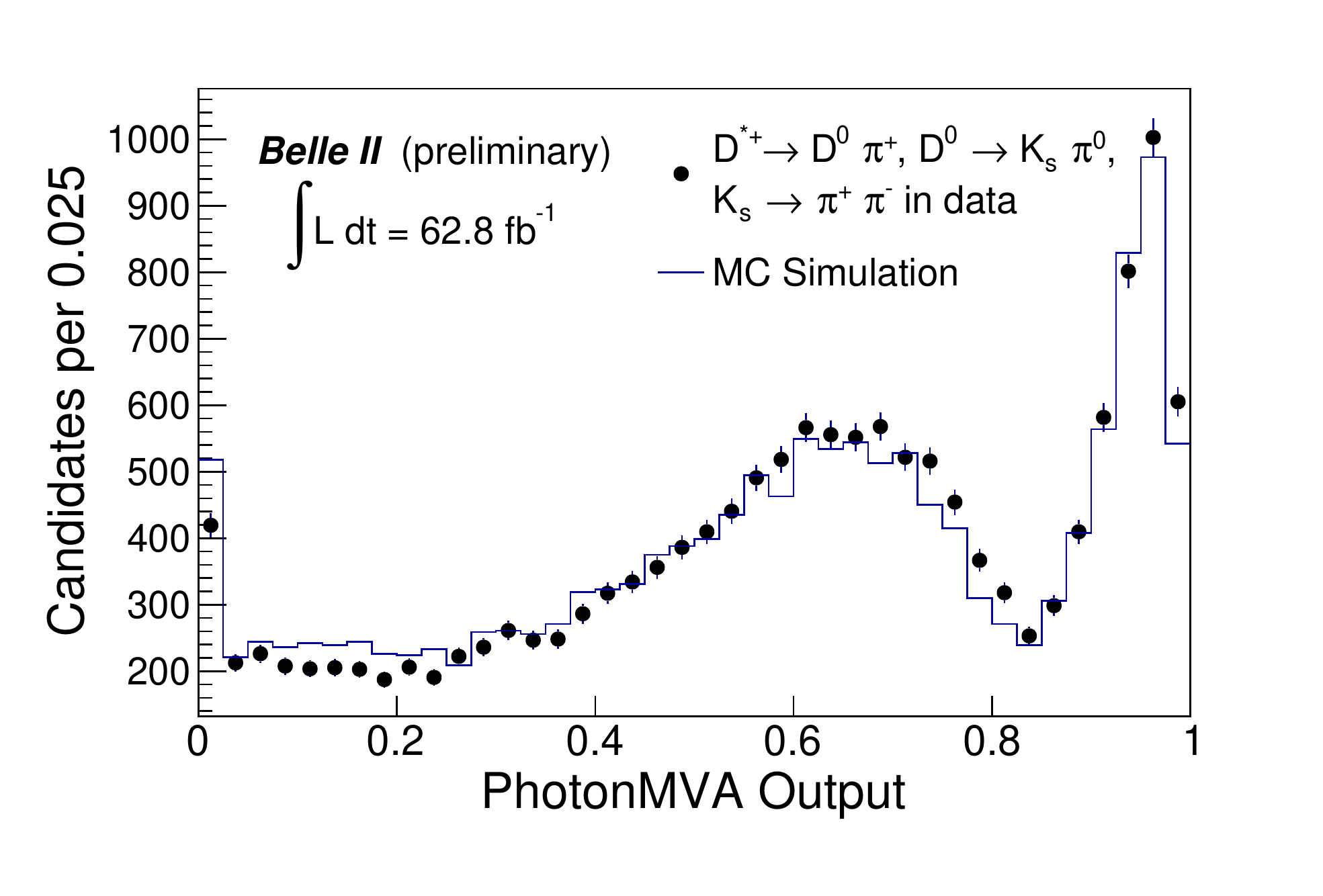}
    \vspace{-0.8cm}
    \caption{The output of the FBDT classifier for photons in $\Dstarp \to \APDzero \pi^+$, $\APDzero \to \KS \pi^0$, $\KS \to \pi^+ \pi^-$ decays.}
    \label{fig:calib_photonmva}
\end{figure}

The final-state reconstruction is performed with and without the \texttt{photonMVA} requirement. The signal yields and backgrounds are determined using fits to the $\Delta M$ distribution. The signal and background retention are consistent between MC and experimental data. Requiring the \texttt{photonMVA} to exceed 0.2 reduces the $D^{*+}$ background by 47\% (absolute) and reduces signal reconstruction efficiency by 3.7\% (absolute) compared to the standard photon selection~\cite{manfredi2021measurements}. In addition we compare the \texttt{photonMVA} distribution in the MC and 62.8 \invfb data as shown in Figure~\ref{fig:calib_photonmva} and find that they are in agreement. 

\section{Multivariate background suppression}
\label{sec:csmva}
The $\Bpipi$ decay has large continuum background even relative to other charmless $B$ decay modes because of its small branching ratio. To discriminate against such background, we use the FBDT classifier to combines 28 variables associated with event topology, which are known to provide statistical discrimination between $B$-meson signal and continuum background. We additionally require that these variables have correlations with $\dele$ and $\mbc$ below 5\%. We train the classifier to identify statistically significant signal and background features using unbiased simulated samples. We validate the input and output distributions of the classifier by comparing data with simulation using the control mode $B^{0} \rightarrow \APDzero(\rightarrow K^- \pi^+ \pi^0) \pi^0$. No inconsistency is observed. 

The optimal FBDT threshold is determined by maximizing the figure of merit $S/\sqrt{S + B}$, where $S$ and $B$ are the simulated signal and background yield, respectively, both determined in the signal-enhanced region $5.27 ~< \mbc ~< 5.29 \gevcc$ and $-0.1 ~<\dele~< 0.15 \gev$. The resulting threshold criterion rejects 97.8\% of the background while retaining 57.9\% of the signal. The signal efficiency after $\Bpipi$ candidate reconstruction and continuum suppression is 21\%. About 1.6\% of selected events have more than one candidate. We choose the candidate with the minimum absolute deviation, $|dM(\pi^0_1)|+|dM(\pi^0_2)|$, of the reconstructed invariant masses from the known value. This is 98\% efficient in selecting the correct $\PBzero$. In addition, we reconstruct the vertex of the accompanying tag-side $B$ meson and identify the flavor using a category-based flavor tagger~\cite{Abudinen:2019hcr}. 

To validate the FBDT, 9.2 \invfb of off-resonance data is used as it contains only continuum events. We apply the same $\Bpipi$ selection as discussed previously except for the requirements $M_{bc} > 5.2 \gevcc$ and $|\dele| < 0.5 \gev$ imposed to retain as many continuum events as possible. We impose the FBDT requirement and determine the continuum rejection to be 97.4\%, which agrees with MC expectation.

Backgrounds due to non-signal $B$ decays are denoted as $\BB$. All the $\BB$ background effectively consists of $B^+ \to \rho^+ \pi^0$ decays where the charged pion from the subsequent $\rho^+ \to \pi^+ \pi^0$ is unreconstructed, and $\PBzero \to K^0_S(\to \pi^0 \pi^0) \pi^0$ decays where one of the $\pi^0$ is unreconstructed. This background peaks at similar values of $\mbc$ and $T_{c}$ but has $\dele$ shifted to negative values due to missing energy from the unreconstructed particle. Since the topology of $\BB$ events is similar to the signal mode, the FBDT removes approximately the same fraction of $\BB$ and signal events. 

\section{Determination of signal yields}
\label{sec:fitting}

Signal yields are determined with a three-dimensional ($\mbc, \dele$, $T_{c}$) simultaneous unbinned maximum likelihood fit for each $b$ flavor, $q$, in 8 bins of the dilution factor, $r$. $T_{c}$ is the log transform of the continuum suppression FBDT variable and is used to transform the FBDT output into a Gaussian-like shape. By convention $q = +1$ tags a $\PBzero$ while $q = -1$ tags a $\APBzero$. The dilution factor is determined by a category-based flavor tagger~\cite{Abudinen:2019hcr} with $r = 0$ meaning no flavor discrimination between $\PBzero$ and $\APBzero$ and $r = 1$ meaning unambiguous flavor assignment. Bins are spaced so that each bin has an approximately equal number of candidates. Fit models are determined empirically from simulation, with shifts of peak positions in $\dele$ determined in the $B^{0} \rightarrow \APDzero(\rightarrow K^- \pi^+ \pi^0) \pi^0$ control mode. A negative shift in $\dele$ relative to simulated data is observed. The value of the shift favored by control data is determined using a likelihood-ratio test to be $-10$ MeV. This is included in the $\dele$ fit model for the signal and $\BB$ component. 

To validate the MC $q \cdot r$ distributions, we compare the $q \cdot r$ fractions in experimental and simulated data restricted to a continuum-enriched sideband defined as $5.20 < \mbc < 5.26 \gevcc$ where continuum events are dominant. The continuum suppression requirement is removed to have a sufficient sample size. The results show excellent agreement. 

For all components, $T_{c}$ is modelled using the sum of a Gaussian and a bifurcated Gaussian with a common mean to avoid peak splitting. For the signal and $\BB$ components, the correlation between $\mbc$ and $\dele$ is taken into account with a two-dimensional kernel density estimation. We employ a data-driven method to determine the parameters of the continuum background probability density function (PDF) by fitting to the sideband region defined as $5.24 < \mbc < 5.27 \gevcc$ and $0.1 < \dele < 0.3 \gev$. The range is limited since selections too far from the signal region may accept candidates too far kinematically from the signal region. The $\BB$ background has a negligible contribution as within both ranges, only 0.07\% survive and we expect $\BB$ to be only a tenth of the total background. All signal and \BB PDF parameters are fixed from simulated data while all continuum PDF parameters are fixed from experimental sideband data. The PDF parameters are identical for all $q.r$ bins. The signal, continuum and \BB bin fractions are fixed while the yields are allowed to float. The full PDF for the signal component is given by
\begin{equation}
    P^s_{i}(\mbc,\dele, T_{c},q) = [1-q \times \Delta w_{i} + q(1-2w_{i}) \times (1-2\chi_d)\mathcal{A}_{CP}]P^s(\mbc,\dele, T_c),
\end{equation}
where $q$ is determined for the $i^{th}$ bin of the data set, $P^s(\mbc,\dele, T_c)$ is the signal PDF in $\mbc$, $\dele$, and $T_c$, $\mathcal{A}_{CP}$ is the direct ${\it CP}$ violation parameter, $\chi_d = 0.1875 \pm 0.0017 $ is the time-integrated $\PBzero \APBzero$-mixing asymmetry, $w_i$ is the wrong-tag fraction, and $\Delta w_i$ is the difference in wrong tag fraction between positive and negative b-flavor tags for bin $i$. The values of $w_i$, and $\Delta w_i$ are all fixed from MC. The $\mathcal{A}_{CP}$ parameter is allowed to float but is not reported due to limited statistics. 

\section{Systematic uncertainties}
\label{sec:systematics}
With the current 62.8 \invfb data set the dominant uncertainty is statistical. As more experimental data are collected, the systematic uncertainties will become more important. At this stage, we only examine the largest sources of systematic uncertainty. We assume the sources to be independent and add in quadrature the corresponding uncertainties. Table~\ref{table:uncertainty} summarizes the systematic uncertainty. 

\subsection{$\pi^0$ reconstruction efficiency}
The systematic uncertainty associated with possible data-simulation discrepancies is studied using the decays $\PBzero \to D^{*-} (\to \APDzero (\to K^+ \pi^- \pi^0) \pi^-) \pi^+$ and $\PBzero \to D^{*-} (\to \APDzero (\to K^+ \pi^-) \pi^-) \pi^+$ where the selection of charged particles is identical and the $\pi^0$ selection is the same as the signal mode. The signal yields of the two control channels are used to determine the $\pi^0$ reconstruction efficiency

\begin{equation}
    \epsilon^{\pi^{0}} = \frac{N(K^- \pi^+ \pi^0)}{N(K^- \pi^+)}\cdot \frac{\mathcal{B}(\APDzero \to K^+ \pi^-)}{\mathcal{B}(\APDzero \to K^+ \pi^- \pi^0) \cdot \mathcal{B}(\pi^0 \to \gamma \gamma)}
\end{equation}
We compare the yields obtained from fits to the $\dele$ distribution of reconstructed $B$ candidates and obtain an efficiency $\epsilon^{\pi^0}_{\rm data}$ in data that agrees with the value observed in simulation within a 10\% uncertainty, which is used as a systematic uncertainty. The uncertainties on the efficiencies for two $\pi^0$s are completely correlated and hence our total systematic uncertainty is 20\%. 

\subsection{Number of $\BB$}
The calculation of the branching fraction uses the number of $\PBzero\APBzero$ pairs, which is computed using the integrated luminosity, the $e^+e^- \to \Y4S$ cross-section~\cite{BaBar:2014omp}, and the known value of the branching fraction of $\Y4S \to \PBzero\APBzero$. We assign a systematic uncertainty of 1.34\%, which includes the uncertainties on the above quantities and uncertainties associated with the beam-energy spread and potential offset of the c.m. energy.

\subsection{Continuum PDF modeling}
The continuum is modelled with an ARGUS function for $\mbc$, a first order Chebyshev function for $\dele$, and the sum of a Gaussian and bifurcated Gaussian for $T_{c}$. The continuum background uses a total of 8 parameters, which are fixed from a fit to the sideband. The sideband for $\mbc$ is defined as $0.1< \dele <0.2$ GeV while the sideband for $\dele$ and $T_{c}$ is defined as $5.26<\mbc<5.27$ GeV/c$^2$. We vary the eight parameters of the sideband-based modeling of continuum accounting for correlation and obtain a 10\% systematic uncertainty. 

\begin{table}[!h]
\begin{tabular}{|c|c|l}
\cline{1-2}
\textbf{Source}                & \textbf{Systematic Uncertainty (\%)} &  \\ \cline{1-2}
$\pi^0$ efficiency             & 20.0                                 &  \\ \cline{1-2}
$N(\BB)$ & 1.34                                 &  \\ \cline{1-2}
Continuum PDF                  & 10.0                                 &  \\ \cline{1-2}
Total                          & 22.4                              &  \\ \cline{1-2}
\end{tabular}
\caption{Major systematic uncertainties where the total is calculated by adding all the systematic uncertainties in quadrature.}
\label{table:uncertainty}
\end{table}

\section{Determination of branching ratio}

We determine the branching fraction as 
\begin{equation}
    \mathcal{B} = \frac{N}{\epsilon \times 2 \times N_{\BB}} 
    \label{eq:BR}
\end{equation}
where $N$ is the signal yield obtained from the fits, $\epsilon$ is the reconstruction and selection efficiency, and $N_{\BB}$ is the number of $\PB\bar{\PBzero}$ pairs. The number of $\PBzero \bar{\PBzero}$ pairs is obtained from the product of the measured integrated luminosity, the $e^+e^- \to \Upsilon(4S)$ cross section ($1.110 \pm 0.008$ nb)~\cite{BaBar:2014omp}, and the $\Upsilon (4S) \to \B\PBzero$ branching fraction ($f^{00} = 0.487 \pm 0.010 \pm 0.008$)~\cite{BaBar:2005uwr}. 

\begin{figure}[htp]
    \vspace{-0.1cm}
    \includegraphics[width=0.6\textwidth]{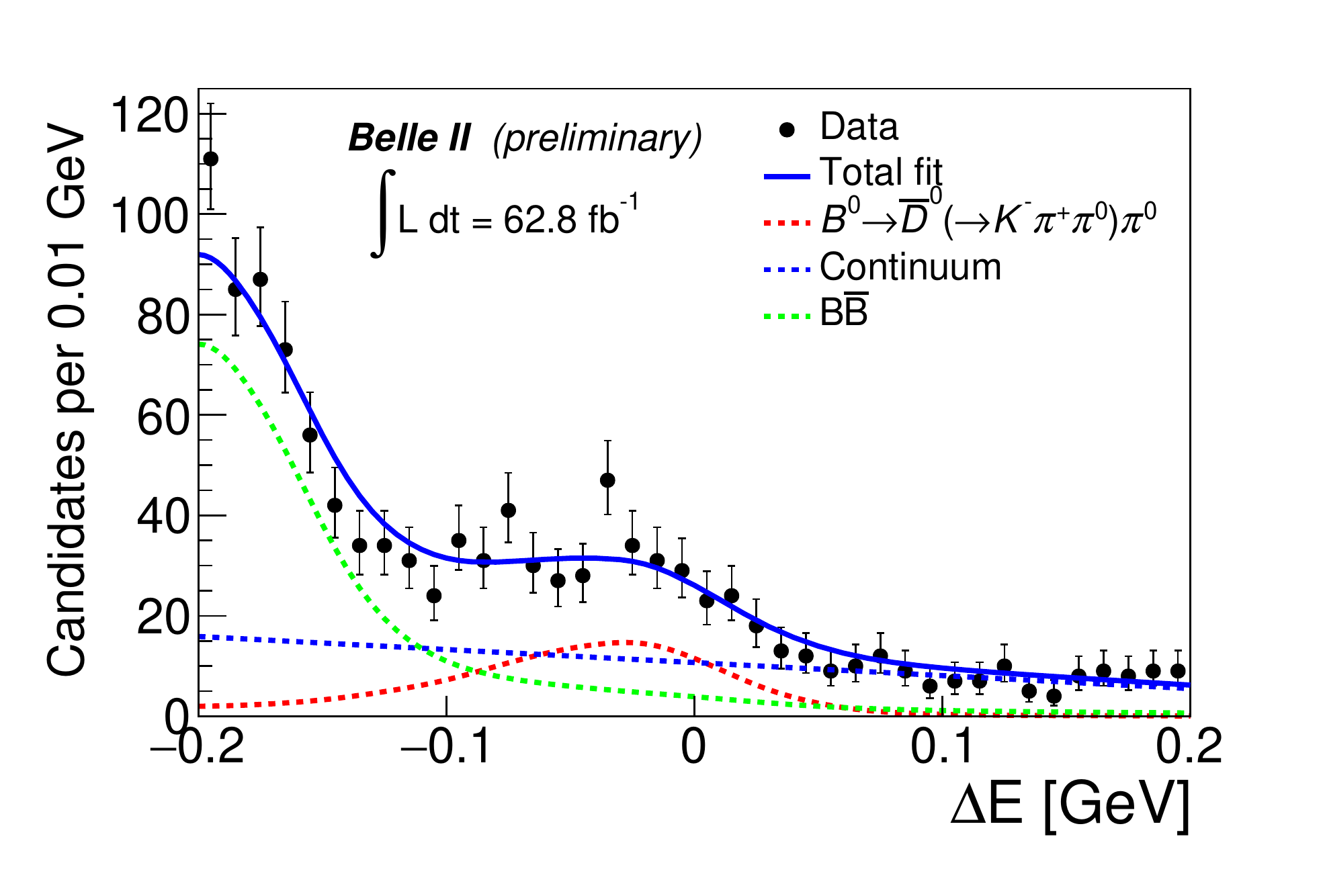}      
    \includegraphics[width=0.6\textwidth]{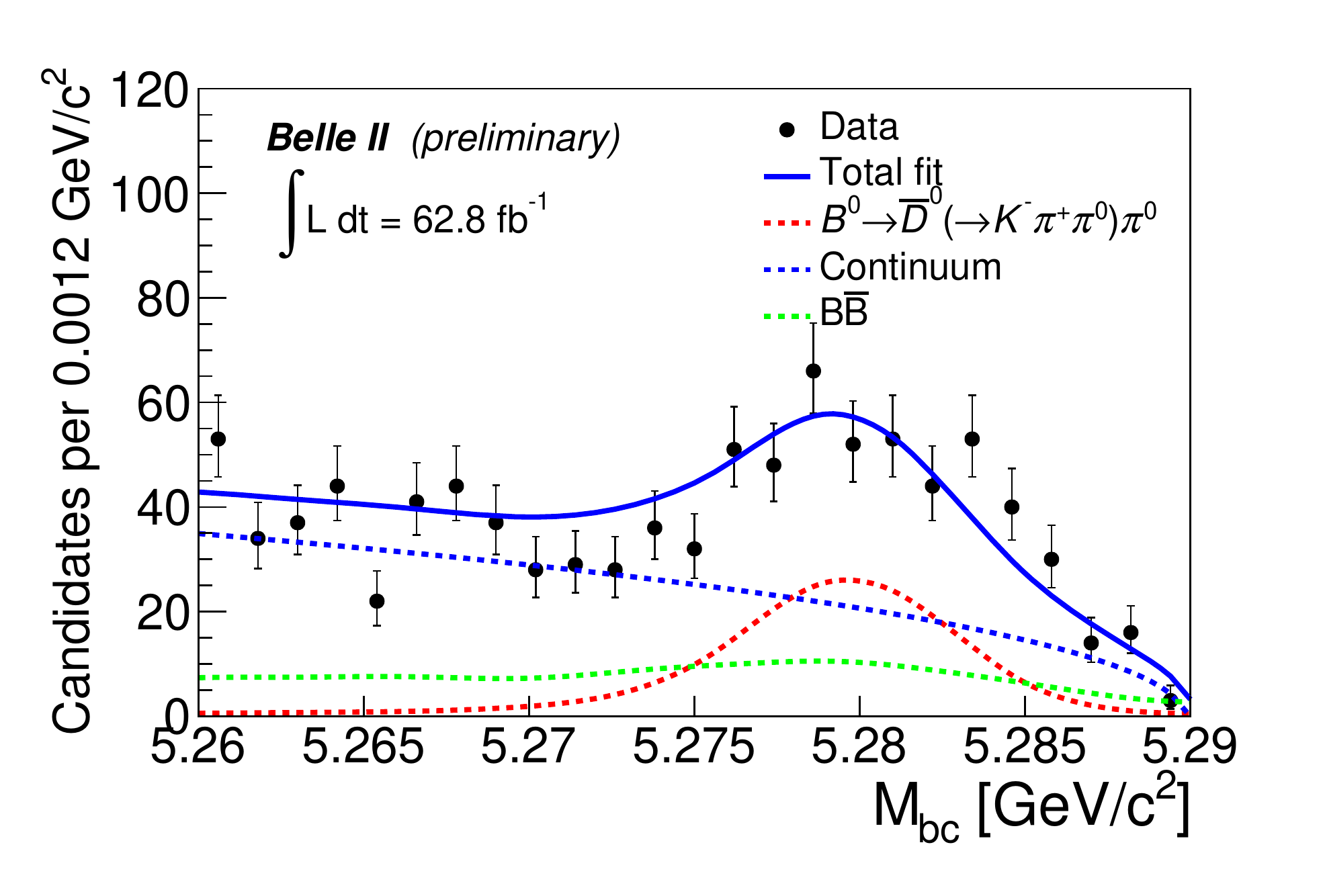}    
    \includegraphics[width=0.6\textwidth]{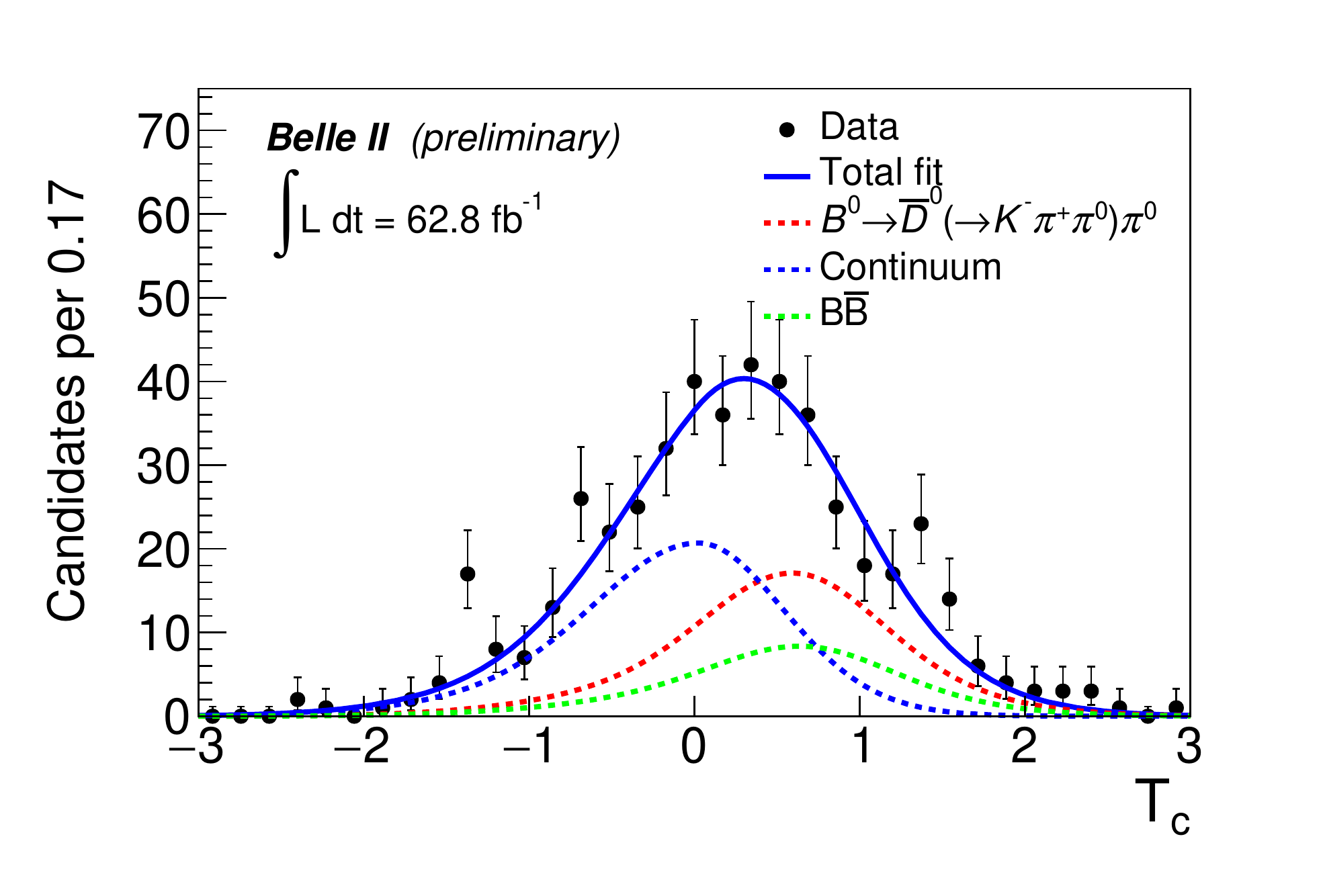}    
    \vspace{-0.2cm}
    \caption{Distributions of $\dele$ (top), $\mbc$ (middle), and $T_{c}$ (bottom) for $B^{0} \rightarrow \APDzero(\rightarrow K^- \pi^+ \pi^0) \pi^0$ reconstructed in 2019–2020 Belle II data. The distributions are shown in signal-enriched regions of $5.275 < \mbc <5.285 \gevcc$ and $-1 < T_{c} < 2$ for $\dele$, $-0.1 < \dele < 0.05 \gev$ and $-1 < T_{c} < 2$ for $\mbc$ and $5.275 < \mbc < 5.285 \gevcc$ and $-0.1 < \dele < 0.05 \gev$ for $T_{c}$. Fit projections are overlaid.}
    \label{fig:controlMORIONDfit}
\end{figure}

\begin{figure}[htp]
    \vspace{-0.1cm}
    \includegraphics[width=0.6\textwidth]{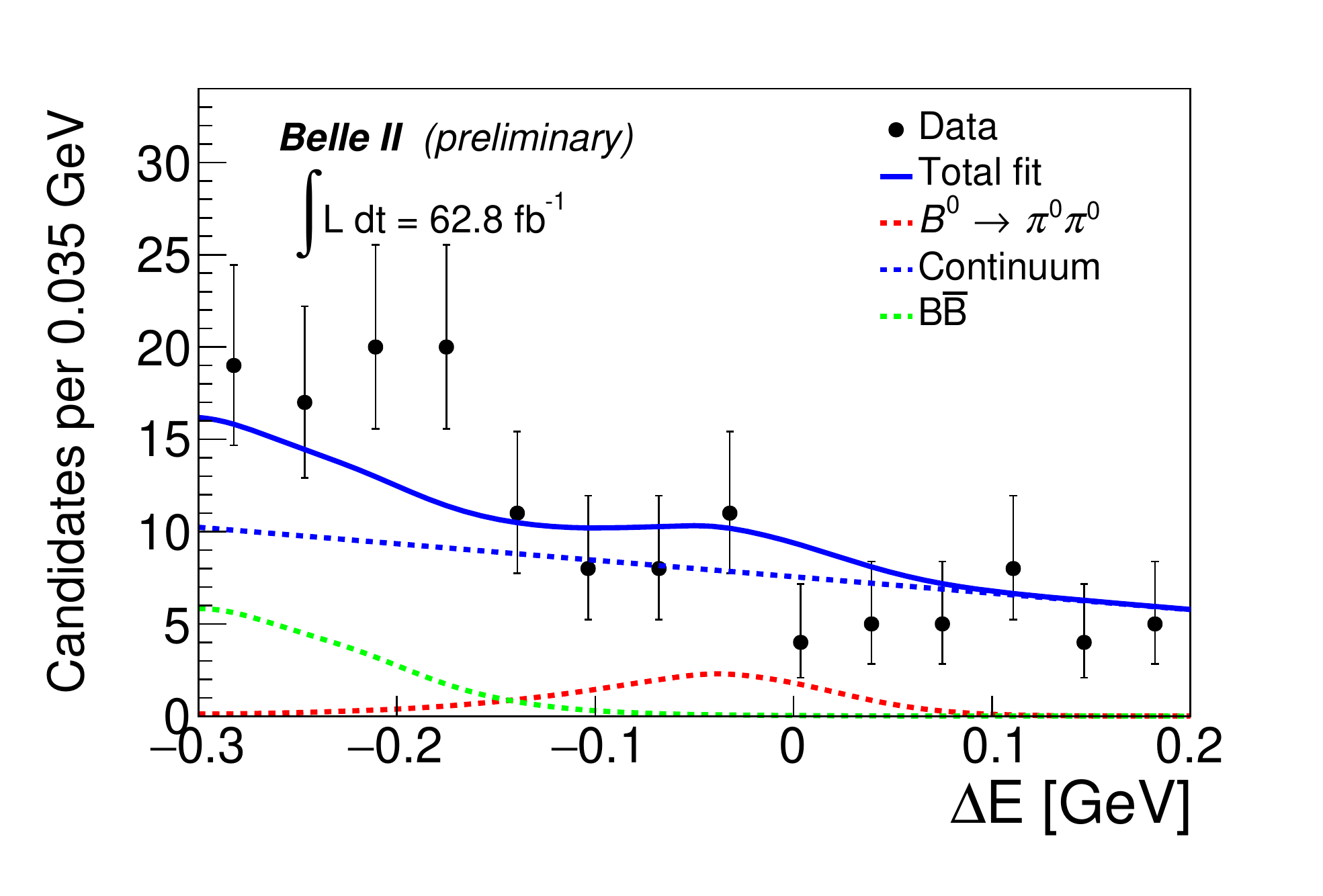}  
    \includegraphics[width=0.6\textwidth]{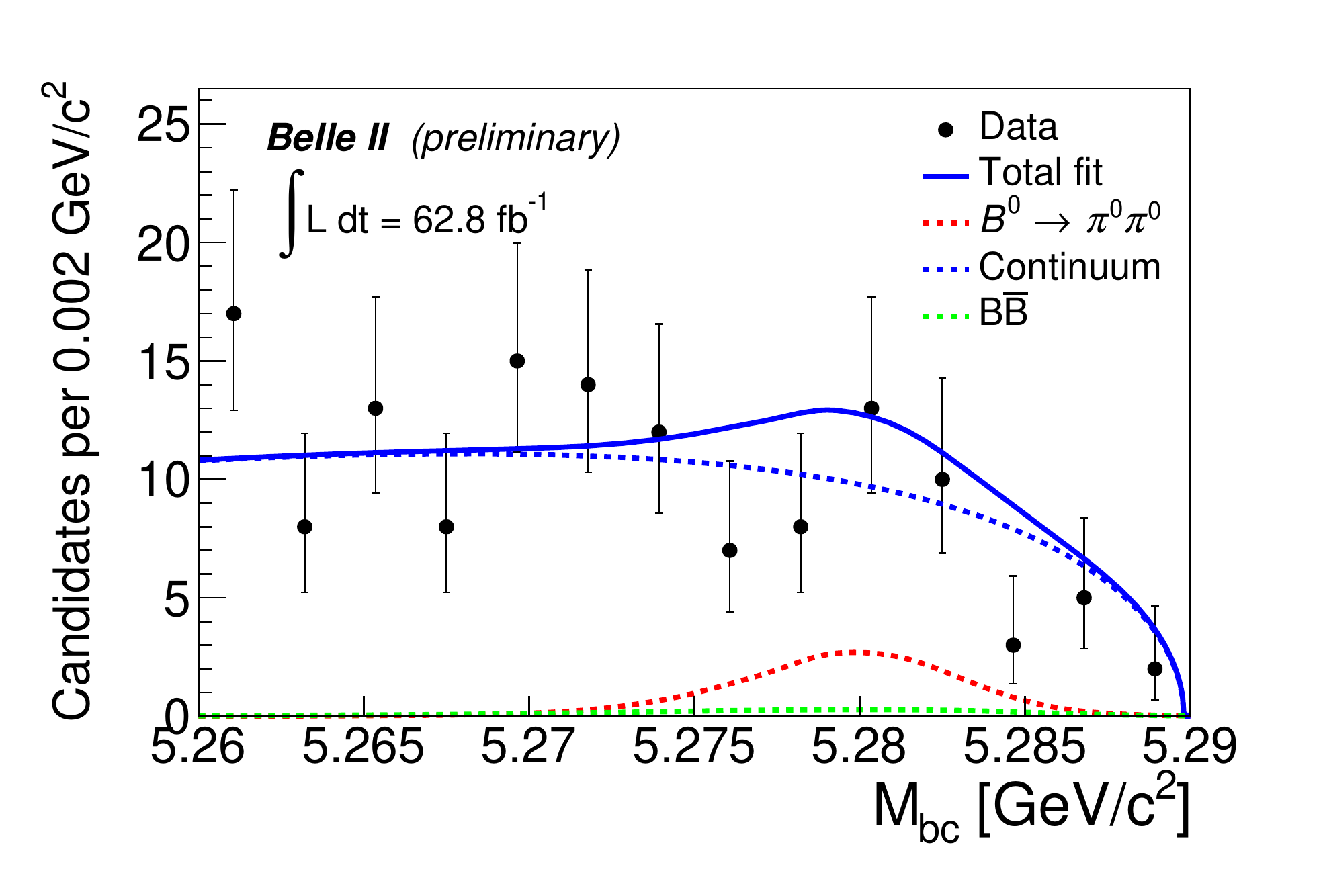}
    \includegraphics[width=0.6\textwidth]{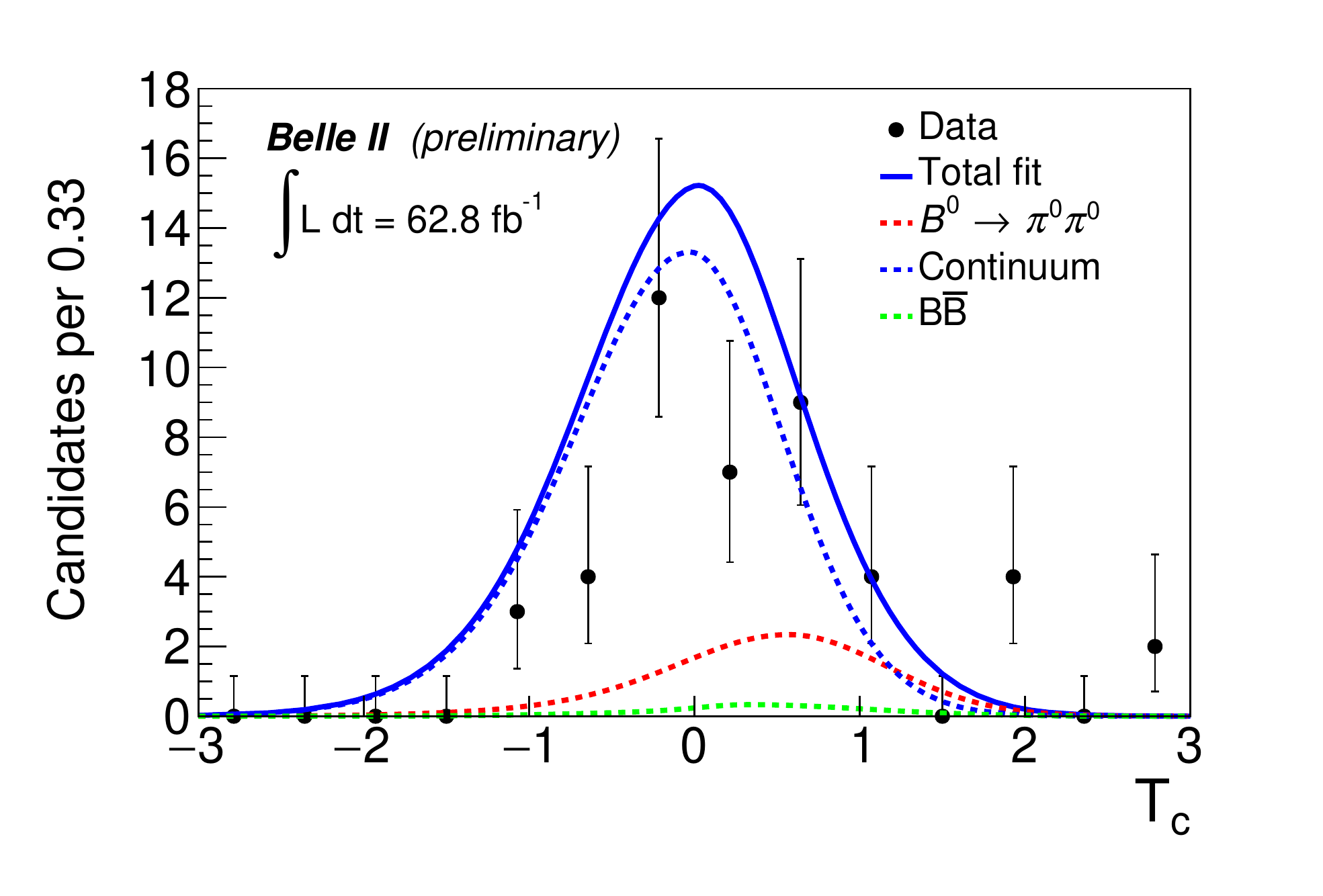}    
    \vspace{-0.2cm}
    \caption{Distributions of $\dele$ (top), $\mbc$ (middle), and $T_{c}$ (bottom) for $\Bpipi$ reconstructed in 2019–2020 Belle II data. The distributions are shown in signal-enriched regions of $5.275 < \mbc <5.285 \gevcc$ and $-1 < T_{c} < 2$ for $\dele$, $-0.1 < \dele < 0.05 \gev$ and $-1 < T_{c} < 2$ for $\mbc$ and $5.275 < \mbc < 5.285 \gevcc$ and $-0.1 < \dele < 0.05 \gev$ for $T_{c}$. Fit projections are overlaid.}
    \label{fig:MORIONDfit}
\end{figure}

Based on simulations, we expect a signal yield of $21.0 \pm 3.4$ events and 373 background events in our selection region. In data, we obtain a signal yield of $14.0^{+6.8}_{-5.6}$ as shown in Figure~\ref{fig:MORIONDfit} with a statistical significance of 3.4 and 403 background events in our selection region. The statistical significance is assessed by comparing the likelihood ratio observed in data with the distribution on a sample of background-only simulated experiments. The branching fraction is calculated to be $\mathcal{B}(\Bpipi) = (0.98^{+0.48}_{-0.39} \pm 0.24) \times 10^{-6}$. The first uncertainties are statistical while the second is systematic. This agrees with the previously measured value, ($1.59 \pm 0.26) \times 10^{-6}$. The result is summarized in Table~\ref{table:sigval}. The yield for the control mode is $295 \pm 31$, as shown in Figure~\ref{fig:controlMORIONDfit}, consistent with 288 expected from simulation. 


\begin{table}[]
\begin{tabular}{cccc}
\hline
\hline
 \qquad Decay  \qquad          & \qquad $\epsilon$ {[}\%{]} \qquad & \qquad Yield  \qquad            & \qquad $\mathcal{B}[10^{-6}]$ \qquad \qquad \\ \hline
\qquad $B^0 \to \pi^0 \pi^0$ & \qquad 21.0                & \qquad $14.0^{+6.8}_{-5.6}$ & $0.98^{+0.48}_{-0.39}$ \qquad \\ \hline \hline
\end{tabular}
\caption{Summary of signal efficiency $\epsilon$, signal yield in 2019-2020 Belle II data and resulting branching fraction. Only the statistical contributions to the uncertainties are given here.}
\label{table:sigval}
\end{table}

\section{Summary}
\label{sec:summary}
We report the first reconstruction of the $B^{0} \to \pi^{0} \pi^{0}$ decay in Belle II using 62.8 \invfb of data. An improved method of selecting photons for signal reconstruction, \texttt{photonMVA}, that utilizes a boosted decision tree rather than rectangular cuts is validated on experimental data. We find a signal yield of $14.0^{+6.8}_{-5.6}$ events, corresponding to a significance of 3.4 standard deviations and determine $\mathcal{B}(\Bpipi) = (0.98^{+0.48}_{-0.39} \pm 0.27) \times 10^{-6}$, where the first uncertainty is statistical and the second, systematic. With much larger data samples that are expected in the near future, Belle II will be able to measure the direct CP violation parameter, $\mathcal{A_{CP}}(B^0\to \pi^0 \pi^0)$ and provide improved constraints on $\alpha / \phi_2$.
\newpage

\section{Acknowledgements}
We thank the SuperKEKB group for the excellent operation of the
accelerator; the KEK cryogenics group for the efficient
operation of the solenoid; the KEK computer group for
on-site computing support; and the raw-data centers at
BNL, DESY, GridKa, IN2P3, and INFN for off-site computing support.
This work was supported by the following funding sources:
Science Committee of the Republic of Armenia Grant No. 20TTCG-1C010;
Australian Research Council and research grant Nos.
DP180102629, 
DP170102389, 
DP170102204, 
DP150103061, 
FT130100303, 
and
FT130100018; 
Austrian Federal Ministry of Education, Science and Research,
Austrian Science Fund No. P 31361-N36, and
Horizon 2020 ERC Starting Grant no. 947006 ``InterLeptons''; 
Natural Sciences and Engineering Research Council of Canada, Compute Canada and CANARIE;
Chinese Academy of Sciences and research grant No. QYZDJ-SSW-SLH011,
National Natural Science Foundation of China and research grant Nos.
11521505,
11575017,
11675166,
11761141009,
11705209,
and
11975076,
LiaoNing Revitalization Talents Program under contract No. XLYC1807135,
Shanghai Municipal Science and Technology Committee under contract No. 19ZR1403000,
Shanghai Pujiang Program under Grant No. 18PJ1401000,
and the CAS Center for Excellence in Particle Physics (CCEPP);
the Ministry of Education, Youth and Sports of the Czech Republic under Contract No.~LTT17020 and 
Charles University grants SVV 260448 and GAUK 404316;
European Research Council, 7th Framework PIEF-GA-2013-622527, 
Horizon 2020 ERC-Advanced Grants No. 267104 and 884719,
Horizon 2020 ERC-Consolidator Grant No. 819127,
Horizon 2020 Marie Sklodowska-Curie grant agreement No. 700525 `NIOBE,' 
and
Horizon 2020 Marie Sklodowska-Curie RISE project JENNIFER2 grant agreement No. 822070 (European grants);
L'Institut National de Physique Nucl\'{e}aire et de Physique des Particules (IN2P3) du CNRS (France);
BMBF, DFG, HGF, MPG, and AvH Foundation (Germany);
Department of Atomic Energy under Project Identification No. RTI 4002 and Department of Science and Technology (India);
Israel Science Foundation grant No. 2476/17,
United States-Israel Binational Science Foundation grant No. 2016113, and
Israel Ministry of Science grant No. 3-16543;
Istituto Nazionale di Fisica Nucleare and the research grants BELLE2;
Japan Society for the Promotion of Science,  Grant-in-Aid for Scientific Research grant Nos.
16H03968, 
16H03993, 
16H06492,
16K05323, 
17H01133, 
17H05405, 
18K03621, 
18H03710, 
18H05226,
19H00682, 
26220706,
and
26400255,
the National Institute of Informatics, and Science Information NETwork 5 (SINET5), 
and
the Ministry of Education, Culture, Sports, Science, and Technology (MEXT) of Japan;  
National Research Foundation (NRF) of Korea Grant Nos.
2016R1\-D1A1B\-01010135,
2016R1\-D1A1B\-02012900,
2018R1\-A2B\-3003643,
2018R1\-A6A1A\-06024970,
2018R1\-D1A1B\-07047294,
2019K1\-A3A7A\-09033840,
and
2019R1\-I1A3A\-01058933,
Radiation Science Research Institute,
Foreign Large-size Research Facility Application Supporting project,
the Global Science Experimental Data Hub Center of the Korea Institute of Science and Technology Information
and
KREONET/GLORIAD;
Universiti Malaya RU grant, Akademi Sains Malaysia and Ministry of Education Malaysia;
Frontiers of Science Program contracts
FOINS-296,
CB-221329,
CB-236394,
CB-254409,
and
CB-180023, and SEP-CINVESTAV research grant 237 (Mexico);
the Polish Ministry of Science and Higher Education and the National Science Center;
the Ministry of Science and Higher Education of the Russian Federation,
Agreement 14.W03.31.0026;
University of Tabuk research grants
S-0256-1438 and S-0280-1439 (Saudi Arabia);
Slovenian Research Agency and research grant Nos.
J1-9124
and
P1-0135; 
Agencia Estatal de Investigacion, Spain grant Nos.
FPA2014-55613-P
and
FPA2017-84445-P,
and
CIDEGENT/2018/020 of Generalitat Valenciana;
Ministry of Science and Technology and research grant Nos.
MOST106-2112-M-002-005-MY3
and
MOST107-2119-M-002-035-MY3, 
and the Ministry of Education (Taiwan);
Thailand Center of Excellence in Physics;
TUBITAK ULAKBIM (Turkey);
Ministry of Education and Science of Ukraine;
the US National Science Foundation and research grant Nos.
PHY-1807007 
and
PHY-1913789, 
and the US Department of Energy and research grant Nos.
DE-AC06-76RLO1830, 
DE-SC0007983, 
DE-SC0009824, 
DE-SC0009973, 
DE-SC0010073, 
DE-SC0010118, 
DE-SC0010504, 
DE-SC0011784, 
DE-SC0012704, 
DE-SC0021274; 
and
the Vietnam Academy of Science and Technology (VAST) under grant DL0000.05/21-23.

\newpage 

\bibliography{belle2}
\bibliographystyle{belle2-note}

\end{document}